\documentclass[prd,twocolumn,preprintnumbers,amsmath,amssymb,nofootinbib]{revtex4}

\usepackage{amsmath}
\usepackage{graphicx}
\usepackage{xspace}
\usepackage{color}
\usepackage{url}
\usepackage[utf8]{inputenc}
\usepackage{epstopdf}
\usepackage{hyperref}
\usepackage{ragged2e}
\usepackage[T1]{fontenc}

\newcommand{\bea}{\begin{eqnarray}}
\newcommand{\eea}{\end{eqnarray}}

\begin{document}
\preprint{CERN-TH-2021-053, PSI-PR-21-05, ZU-TH 16/21}

\title{Addendum to: Combined Constraints on First Generation Leptoquarks}

\author{Andreas Crivellin}
\email{andreas.crivellin@cern.ch}
\affiliation{CERN Theory Division, CH--1211 Geneva 23, Switzerland}
\affiliation{Physik-Institut, Universit\"at Z\"urich,
	Winterthurerstrasse 190, CH-8057 Z\"urich, Switzerland}
\affiliation{Paul Scherrer Institut, CH--5232 Villigen PSI, Switzerland}

\author{Luc Schnell}
\email{luschnel@student.ethz.ch}
\affiliation{Laboratoire de Physique Th\'eorique et Hautes \'Energies, LPTHE, Sorbonne Universit\'e, CNRS, 4 place Jussieu, FR-75252 Paris Cedex 05, France}
\affiliation{Departement Physik, ETH Zürich, Otto-Stern-Weg 1, CH-8093 Zürich, Switzerland}
\affiliation{D\'epartement de Physique, \'Ecole Polytechnique, Route de Saclay, FR-91128 Palaiseau Cedex, France}

\begin{abstract}
{In this addendum to Ref.~\cite{Crivellin:2021egp} we discuss the implications of the recent CMS analysis of lepton flavour universality violation in non-resonant di-lepton pairs for first generation leptoquarks. As CMS finds more electron events than expected from background, this analysis prefers the LQ representations $\tilde{S}_1, S_2, S_3, \tilde{V}_1, V_2\,(\kappa_2^{RL}\ne 0)$ and $V_3$ which lead to constructive interference with the SM. In principle the excess could also be (partially) explained by the representations $\tilde{S}_2, V_1\,(\kappa_1^R\ne 0), V_2\,(\kappa_2^{LR}\ne 0), \tilde{V}_2$ which are interfering destructively, as this would still lead to the right effect in bins with high invariant mass where the new physics contribution dominates. However, in these cases large couplings would be required which are excluded by other observables. The representations {$S_1, V_1\, (\kappa_1^{L} \ne 0)$} cannot improve the fit to the CMS data compared to the SM.}
\end{abstract}

\keywords{Leptoquarks, LHC, non-resonant di-leptons}

\maketitle

\section{Introduction}

Leptoquarks are very well motivated by the so-called ``flavor anomalies'', i.e. the discrepancies between measurements and the SM predictions which point towards lepton flavor universality (LFU) violating new physics (NP) in $R(D^{(*)})$~\cite{Lees:2012xj,Lees:2013uzd,Aaij:2015yra,Aaij:2017deq,Aaij:2017uff,Abdesselam:2019dgh}, $b\to s\ell^{+}\ell^{-}$~\cite{CMS:2014xfa,Aaij:2015oid,Abdesselam:2016llu,Aaij:2017vbb,Aaij:2019wad,Aaij:2020nrf,Aaij:2021vac} and in the anomalous magnetic moment (AMM) of the muon ($a_\mu$)~\cite{Bennett:2006fi,Abi:2021gix}, with a significance of~$>\!3\,\sigma$~\cite{Amhis:2016xyh,Murgui:2019czp,Shi:2019gxi,Blanke:2019qrx,Kumbhakar:2019avh}, $>\!5\sigma$~\cite{Capdevila:2017bsm, Altmannshofer:2017yso,Alguero:2019ptt,Alok:2019ufo,Ciuchini:2019usw,Aebischer:2019mlg, Arbey:2019duh,Kumar:2019nfv} and \mbox{$4.2\,\sigma$}~\cite{Aoyama:2020ynm}, respectively. In this context, it has been shown that LQs can explain $b\to s\ell^+\ell^-$ data~\cite{Alonso:2015sja, Calibbi:2015kma, Hiller:2016kry, Bhattacharya:2016mcc, Buttazzo:2017ixm, Barbieri:2015yvd, Barbieri:2016las, Calibbi:2017qbu, Crivellin:2017dsk, Bordone:2018nbg, Kumar:2018kmr, Crivellin:2018yvo, Crivellin:2019szf, Cornella:2019hct, Bordone:2019uzc, Bernigaud:2019bfy,Aebischer:2018acj,Fuentes-Martin:2019ign,Popov:2019tyc,Fajfer:2015ycq,Blanke:2018sro,deMedeirosVarzielas:2019lgb,Varzielas:2015iva,Crivellin:2019dwb,Saad:2020ihm,Saad:2020ucl,Gherardi:2020qhc,DaRold:2020bib}, $R(D^{(*)})$~\cite{Alonso:2015sja, Calibbi:2015kma, Fajfer:2015ycq, Bhattacharya:2016mcc, Buttazzo:2017ixm, Barbieri:2015yvd, Barbieri:2016las, Calibbi:2017qbu, Bordone:2017bld, Bordone:2018nbg, Kumar:2018kmr, Biswas:2018snp, Crivellin:2018yvo, Blanke:2018sro, Heeck:2018ntp,deMedeirosVarzielas:2019lgb, Cornella:2019hct, Bordone:2019uzc,Sahoo:2015wya, Chen:2016dip, Dey:2017ede, Becirevic:2017jtw, Chauhan:2017ndd, Becirevic:2018afm, Popov:2019tyc,Fajfer:2012jt, Deshpande:2012rr, Freytsis:2015qca, Bauer:2015knc, Li:2016vvp, Zhu:2016xdg, Popov:2016fzr, Deshpand:2016cpw, Becirevic:2016oho, Cai:2017wry, Altmannshofer:2017poe, Kamali:2018fhr, Mandal:2018kau, Azatov:2018knx, Wei:2018vmk, Angelescu:2018tyl, Kim:2018oih, Aydemir:2019ynb, Crivellin:2019qnh, Yan:2019hpm,Crivellin:2017zlb, Marzocca:2018wcf, Bigaran:2019bqv,Crivellin:2019dwb,Saad:2020ihm,Dev:2020qet,Saad:2020ucl,Altmannshofer:2020axr,Fuentes-Martin:2020bnh,Gherardi:2020qhc,DaRold:2020bib} and/or $a_\mu$~\cite{Bauer:2015knc,Djouadi:1989md, Chakraverty:2001yg,Cheung:2001ip,Popov:2016fzr,Chen:2016dip,Biggio:2016wyy,Davidson:1993qk,Couture:1995he,Mahanta:2001yc,Queiroz:2014pra,ColuccioLeskow:2016dox,Chen:2017hir,Das:2016vkr,Crivellin:2017zlb,Cai:2017wry,Crivellin:2018qmi,Kowalska:2018ulj,Dorsner:2019itg,Crivellin:2019dwb,DelleRose:2020qak,Saad:2020ihm,Bigaran:2020jil,Dorsner:2020aaz,Fuentes-Martin:2020bnh,Gherardi:2020qhc,Babu:2020hun,Crivellin:2020tsz}. They have been studied in direct LHC searches~\cite{Kramer:1997hh,Kramer:2004df,Faroughy:2016osc,Greljo:2017vvb, Dorsner:2017ufx, Cerri:2018ypt, Bandyopadhyay:2018syt, Hiller:2018wbv,Faber:2018afz,Schmaltz:2018nls,Chandak:2019iwj,Allanach:2019zfr, Buonocore:2020erb,Borschensky:2020hot}, leptonic observables~\cite{Crivellin:2020mjs} and oblique electroweak (EW) parameters as well as Higgs couplings to gauge bosons~\cite{Keith:1997fv,Dorsner:2016wpm,Bhaskar:2020kdr,Zhang:2019jwp,Gherardi:2020det,Crivellin:2020ukd}. Furthermore, if the LQs couple to first generation fermions particularly many low energy precision probes can be affected~\cite{Shanker:1981mj,Shanker:1982nd,Leurer:1993em,Leurer:1993qx,Davidson:1993qk}, including the so-called ``Cabibbo-Angle Anomaly''~\cite{Grossman:2019bzp,Seng:2020wjq,Belfatto:2019swo,Coutinho:2019aiy,Crivellin:2020lzu,Capdevila:2020rrl,Crivellin:2020ebi,Kirk:2020wdk,Alok:2020jod,Crivellin:2020oup,Crivellin:2020klg,Crivellin:2021njn,Belfatto:2021jhf,Branco:2021vhs} as well as rare Kaon decays and/or $D^0-\bar D^0$~\cite{Bobeth:2017ecx,Dorsner:2019vgp,Mandal:2019gff}.

In Ref.~\cite{Crivellin:2021egp} we studied the interplay of low and high energy constraints on first generation leptoquarks (LQs). In this addendum we update this analysis by including the recent CMS measurement of lepton flavour universality violation (LFUV) in non-resonant di-leptons~\cite{Sirunyan:2021khd}. The CMS data points towards constructively interfering new physics in the electron channel which can improve the fit compared to the SM by more than $3\,\sigma$~\cite{Crivellin:2021rbf}.

\section{Setup}
\label{sec:setup}
Let us briefly review our conventions. For details, the interested reader is referred to  Ref.~\cite{Crivellin:2021egp}. All 10 possible LQ representations under the SM gauge group are listed in Table~\ref{tab:LQ_representations} with the convention that the electric charge $Q$ is given by $Q=\frac{1}{2}Y+T_{3}$, where $Y$ is the hypercharge and $T_{3}$ the third component of the weak isospin. These representations allow for couplings to SM quarks and leptons as given in Table~\ref{tab:LQ_fermion_coupling}. In the following, we denote the LQ masses according to their representation and use small $m$ for the scalar LQs and capital $M$ for the vector LQs.
\begin{center}
	\begin{table}
		\begin{tabular}{c|ccccc|ccccc}
			Field & $\Phi_{1}$& $\tilde{\Phi}_{1}$ & $\Phi_{2}$ & $\tilde{\Phi}_{2}$ & $\Phi_{3}$ & $V_{1}^{}$ & $\tilde{V}_{1}^{}$ & $V_{2}^{}$ & $\tilde{V}_{2}^{}$ & $V_{3}^{}$\\
			\hline
			$SU(3)_{c}$ & 3 & 3 & 3 & 3 & 3 & 3 & 3 & 3 & 3 & 3\\
			$SU(2)_{L}$ & 1 & 1 & 2 & 2 & 3 & 1 & 1 & 2 & 2 & 3\\
			$U(1)_{Y}$ & $-\frac{2}{3}$ & $-\frac{8}{3}$ & $\frac{7}{3}$ & $\frac{1}{3}$ & $-\frac{2}{3}$ & $\frac{4}{3}$ & $\frac{10}{3}$ & $-\frac{5}{3}$ & $\frac{1}{3}$ & $\frac{4}{3}$
		\end{tabular}
		\caption{The ten possible representations of scalar and vector LQs under the SM gauge group.}
		\label{tab:LQ_representations}
	\end{table}
\end{center}
\begin{table}[t!]
	\renewcommand{\arraystretch}{1.7}
	\begin{tabular}{c|cc}
		{}& $L$&$e$\\
		\hline
		${\bar Q}$& ${\kappa_{1}^L{\gamma _\mu }V_1^{\mu} + \kappa _{3}{\gamma _\mu }\left( {\tau \cdot V_3^\mu } \right)}$&${\lambda_{2}^{LR}{\Phi _2}}$\\
		${\bar d}$& ${\tilde \lambda _{2}\tilde \Phi _2^Ti{\tau _2}}$&${\kappa_{1}^{R}{\gamma _\mu }V_1^{\mu}}$\\
		${\bar u}$& ${\lambda_{2}^{RL}\Phi _2^Ti{\tau _2}}$&${\tilde \kappa_1{\gamma _\mu }\tilde V_1^{\mu }}$\\
		${\bar Q_{}^c}$& ${\lambda_{3}i{\tau _2}{{\left( {\tau \cdot{\Phi _3}} \right)}^\dag } + \lambda_{1}^{L}i{\tau _2}\Phi _1^\dag }$&$\kappa_{2}^{LR}{\gamma _\mu }{V_2^{\mu \dag }}$\\
		${\bar d_{}^c}$& ${\kappa_{2}^{RL}{\gamma _\mu }V_2^{\mu \dag} }$&${\tilde \lambda_{1}\tilde \Phi _1^\dag }$\\
		${\bar u_{}^c}$& $\tilde{\kappa}_{2}{{\gamma _\mu }\tilde V_2^{\mu \dag }}$&${\lambda_{1}^{R}\Phi _1^\dag }$
	\end{tabular}
	\caption{Interaction terms of the LQ representations listed in Table~\ref{tab:LQ_representations}, where $Q$ and $L$ represent the left-handed quark and lepton $SU(2)_{L}$ doublets, $e$, $d$ and $u$ the right-handed $SU(2)_L$ singlets, the superscript $c$ stands for charge conjugation and $\tau_{i}$ are the Pauli matrices.}
	\label{tab:LQ_fermion_coupling}
\end{table}

\section{CMS Analysis of non-resonant di-lepton pairs}
\label{sec:CMS}
\begin{figure*}
\includegraphics[width=0.49\textwidth]{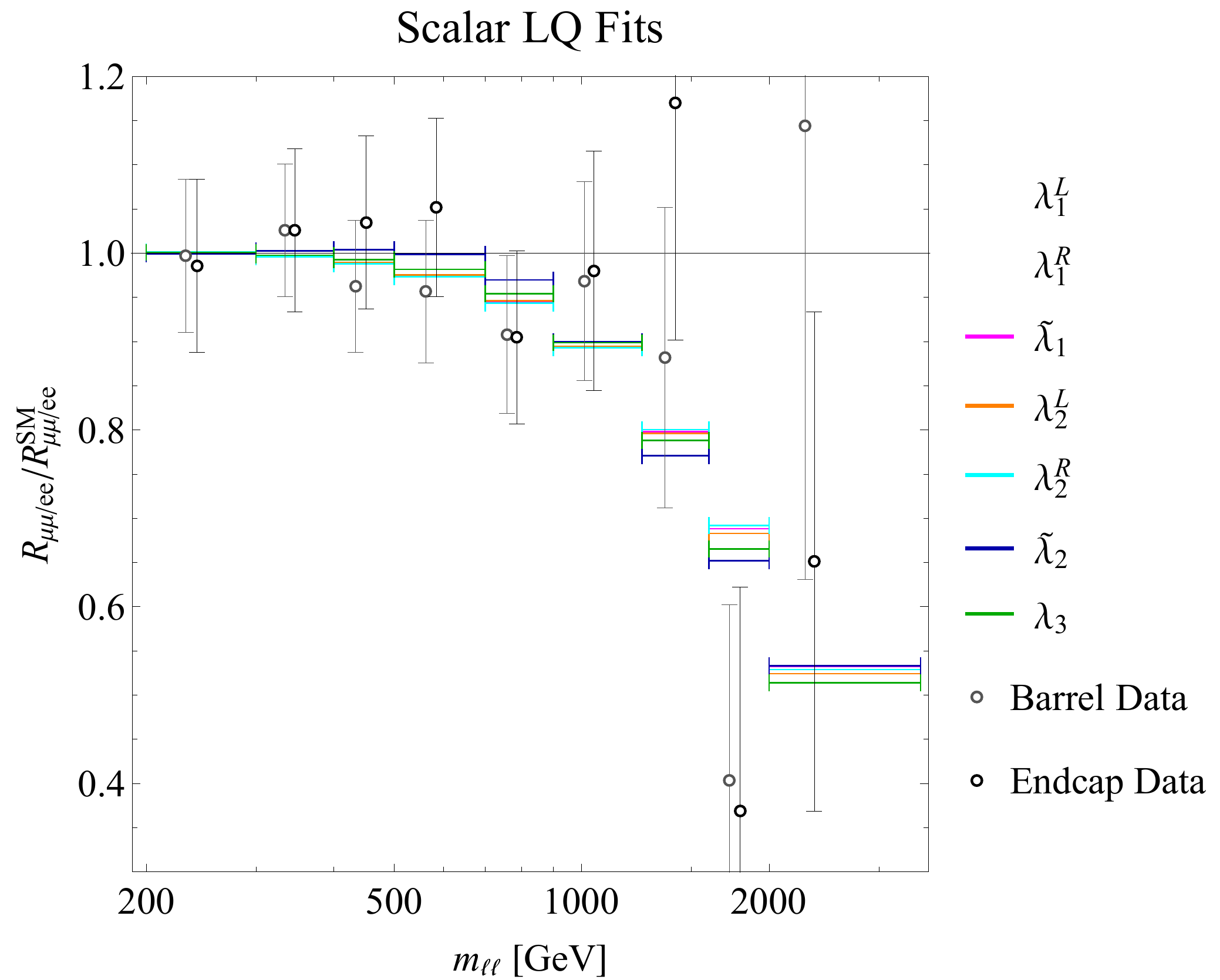}
\includegraphics[width=0.49\textwidth]{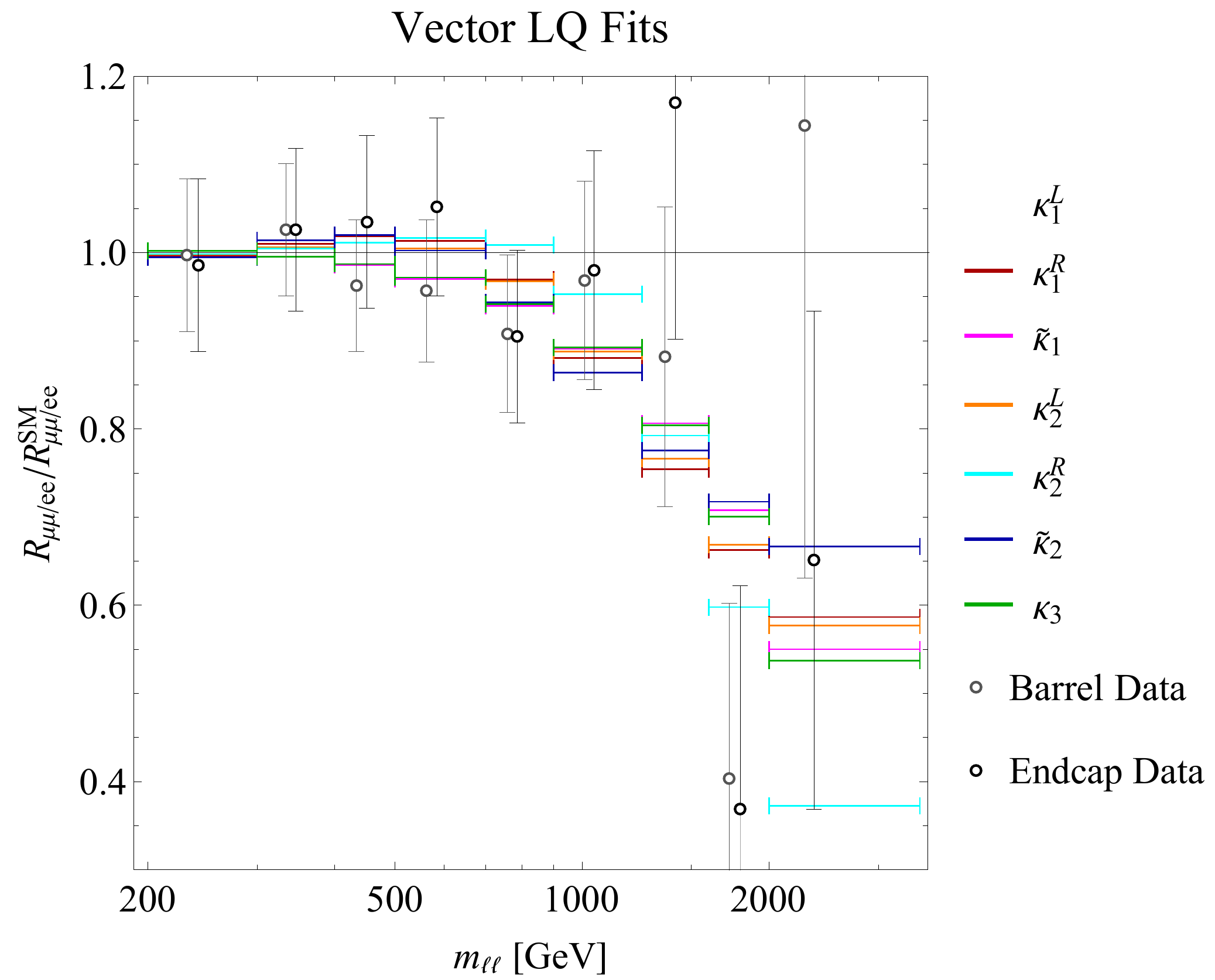}
	\caption{The ratio $R_{\mu\mu/ee}^{\text{Data}}/R_{\mu\mu/ee}^{\text{MC}}$ measured by CMS \cite{Sirunyan:2021khd} in the various bins compared to the $R_{\mu\mu/ee}^{\text{SM+LQ}}/R_{\mu\mu/ee}^{\text{SM}}$ values for the best fit of the various LQ models. The gray data points correspond to measurements where both leptons were detected in the CMS barrel while the black dots correspond to measurements where at least one of the leptons was detected in the CMS endcaps. The parameters of the fit to the different LQ representations are given in Tab.~\ref{tab:Fit}.}
	\label{fig:overview}
\end{figure*}

In Ref.~\cite{Sirunyan:2021khd} CMS presented an analysis of non-resonant high-mass di-lepton events at the LHC. Since in our framework of first generation LQs we only get effects in electrons, we can make use of the ratio
\begin{equation}
    R_{\mu\mu / ee} \equiv \frac{d\sigma(q\overline{q} \to \mu^+\mu^-)/dm_{\mu\mu}}{d\sigma(q\overline{q} \to e^+e^-)/dm_{ee}}\,,
\end{equation}
of differential cross sections, measuring lepton flavour universality (which reduces the uncertainties~\cite{Greljo:2017vvb}), to derive bounds on the LQ masses and couplings. Here, $m_{\ell \ell}$ ($\ell = \mu, e$) is the invariant mass of the lepton pair. In Ref.~\cite{Sirunyan:2021khd} $R_{\mu\mu/ee}$ was measured for nine bins in an invariant mass range between 200 and 3,500~GeV ($R_{\mu\mu/ee}^{\text{Data}}$) and normalized to the same ratio calculated via SM Monte-Carlo routines ($R_{\mu\mu/ee}^{\text{MC}}$). This ratio in turn was normalized to unity in the bin from 200 to 400 GeV in order to correct for differences in acceptance and efficiency between the di-electron and di-muon channels. The resulting data points are shown in Fig.~\ref{fig:overview} where gray (black) represents the cases where none (at least one) of the final state leptons were observed in the detector endcaps.

We calculated the ratio $R_{\mu\mu/ee}^{\text{LQ}}(m, \lambda)/R_{\mu\mu/ee}^{\text{SM}}$ for the different LQ models (at leading order in perturbation theory) using the PDF set NNPDF23LO, also employed e.g. in the ATLAS analysis to generate the signal DY process~\cite{Aad:2020otl}, with the help of the \texttt{Mathematica} package \texttt{ManeParse}~\cite{Clark:2016jgm}. We then integrated this ratio over the invariant mass ranges of the corresponding bins and compared the resulting signal strength to the data. Here a complication arises if the LQ mass is not much higher than the energy of the lepton pair such that the 4-Fermi approximation is no longer appropriate. In this case we replaced the effective interaction by the LQ propagator as described in Sec.~4 in Ref.~\cite{Bessaa:2014jya}. Therefore, $R_{\mu\mu/ee}^{\text{LQ}}$ is dependent both on the LQ mass $m$ and the LQ coupling strength $\lambda$\footnote{Here we consider scalar LQ parameters for simplicity, it works analogously for vector LQs with the parameters $m$, $\lambda$ replaced by $M, \kappa$.}.  We also estimated the relative sensitivities of the CMS detector to the LR/RL vs LL/RR channels, resulting from their different angular distributions, based on the CI limits stated in Ref.~\cite{Sirunyan:2021khd}. We found a small enhancement of 10\% for the LR/RL channels and therefore decided to neglect this effect, obtaining a conservative estimate of the LR/RL LQ contributions in our calculations.

Then we performed a $\chi^2$ statistical analysis with two degrees of freedom, defining 
\begin{equation}
\chi^2(m, \lambda) \equiv \sum_{i = 1, \dots,18}\dfrac{\left(\dfrac{R_{\mu\mu/ee, i}^{\text{Data}}}{R_{\mu\mu/ee, i}^{\text{MC}}} - \dfrac{R_{\mu\mu/ee, i}^{\text{SM+LQ}}(m, \lambda)}{R_{\mu\mu/ee, i}^{\text{SM}}} \right)^2}{\sigma_i^2} \,,
\end{equation}
where $i$ runs over the data points available from the nine bins with and without leptons detected in the CMS endcaps and $\sigma_i$ are the corresponding uncertainties reported in Ref.~\cite{Sirunyan:2021khd}. Minimizing the $\chi^2$ function with respect to $(m, \lambda)$ we find the best fit points $\hat{m}$ and $\hat{\lambda}$ given in Tab.~\ref{tab:Fit}. The corresponding values of $R_{\mu\mu/ee}^{\text{SM+LQ}}/R_{\mu\mu/ee}^{\text{SM}}$ for the different bins are shown in Fig.~\ref{fig:overview}, together with the CMS measurement. 

    	\begin{center}
            	\begin{table*}[t!]
            	    \setlength{\tabcolsep}{0.5em}
            		\begin{tabular}{c|ccccccc|ccccccc}
            			 & $\lambda^L_{1}$ & $\lambda^R_{1}$ & $\tilde{\lambda}_1$ & $\lambda_2^{LR}$ & $\lambda^{RL}_{2}$ & $\tilde{\lambda}_{2}$ & $\lambda_3$ & $\kappa_{1}^{L}$ &$\kappa_{1}^{R}$& $\tilde{\kappa}_{1}$&$\kappa_{2}^{LR}$&$\kappa_{2}^{RL}$& $\tilde{\kappa}_{2}$ & $\kappa_{3}$ \\
            			\hline
            			$\hat{m}$ or $\hat{M}$ [TeV] & $\infty$ & $\infty$ & 3.1 & 2.6 & 3.0 & 1.5 & 2.2 & $\infty$ & 1.0 & 4.3 & 1.0 & 2.5 & 0.6 & 3.9\\
$\hat{\lambda}$ or $\hat{\kappa}$ & 0 & 0 & 0.91 & 0.65 & 0.75 & 0.71 & 0.55 & 0 & 0.45 & 0.61 & 0.35 & 0.60 & 0.35 & 0.40\\
            			$\Delta \chi^2$ &0 & 0 & -10.9 & -10.9 & -10.8 & -11.2 & -11.2 & 0 & -10.5 & -10.5 & -10.8 & -11.1 & -10.0 & -10.6\\
            		\end{tabular}
\caption{Best fit points for the masses ($\hat{m}, \hat{M}$) and couplings ($\hat{\lambda}, \hat{\kappa}$) for the various LQ representations as well as the corresponding minima of $\chi^2$ function with respect to the SM values ($\Delta \chi^2 \equiv \chi^2(\infty, 0) - \chi^2(\hat{m}, \hat{\lambda})$). The corresponding values of $R_{\mu\mu/ee}^{\text{SM+LQ}}/R_{\mu\mu/ee}^{\text{SM}}$ in the different bins are shown in Fig.~\ref{fig:overview}. }
            		\label{tab:Fit}
            	\end{table*}
    	\end{center}
\begin{figure*}
	\begin{center}
		\includegraphics[width=0.45\textwidth]{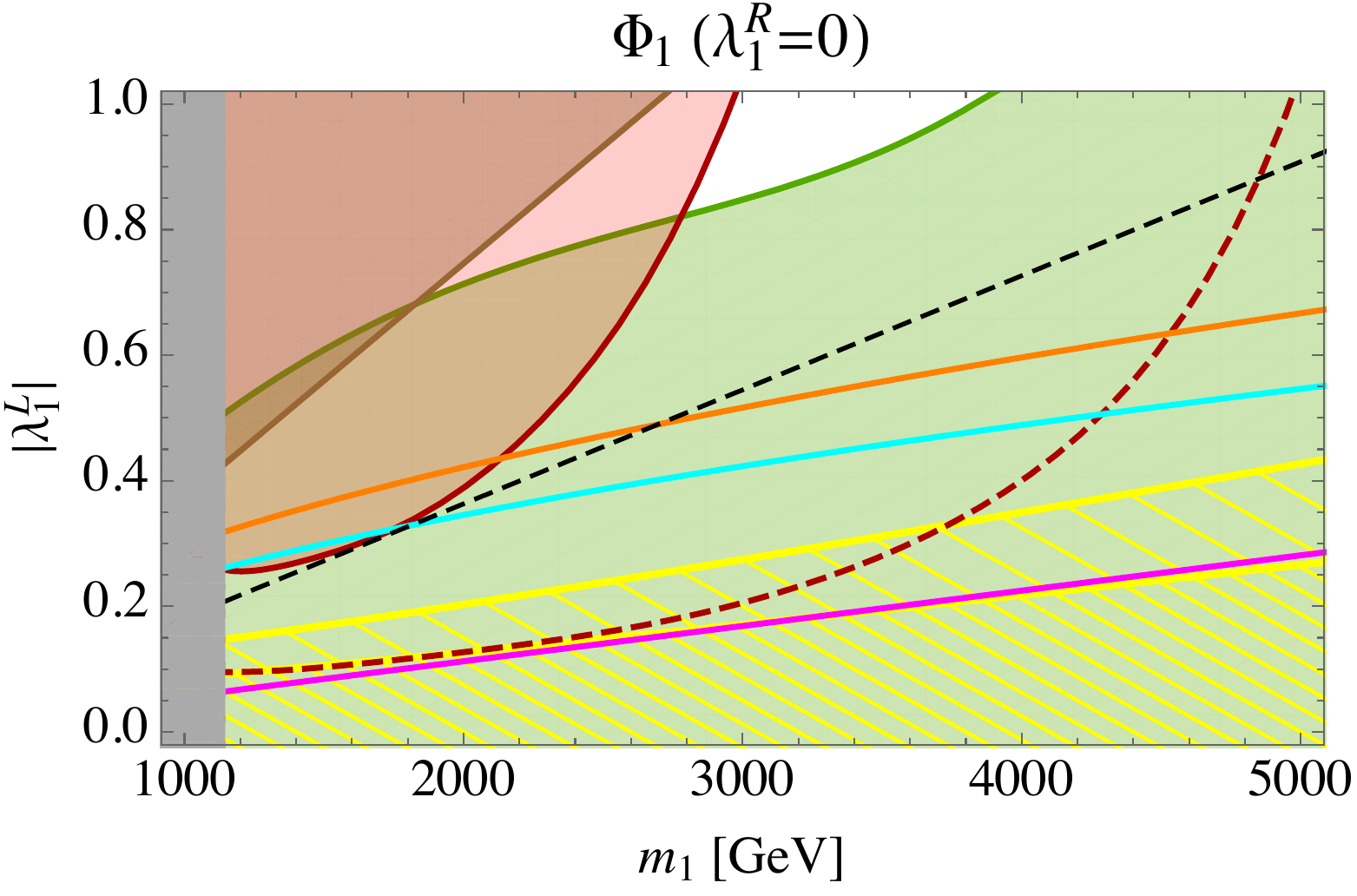}\quad
		\includegraphics[width=0.45\textwidth]{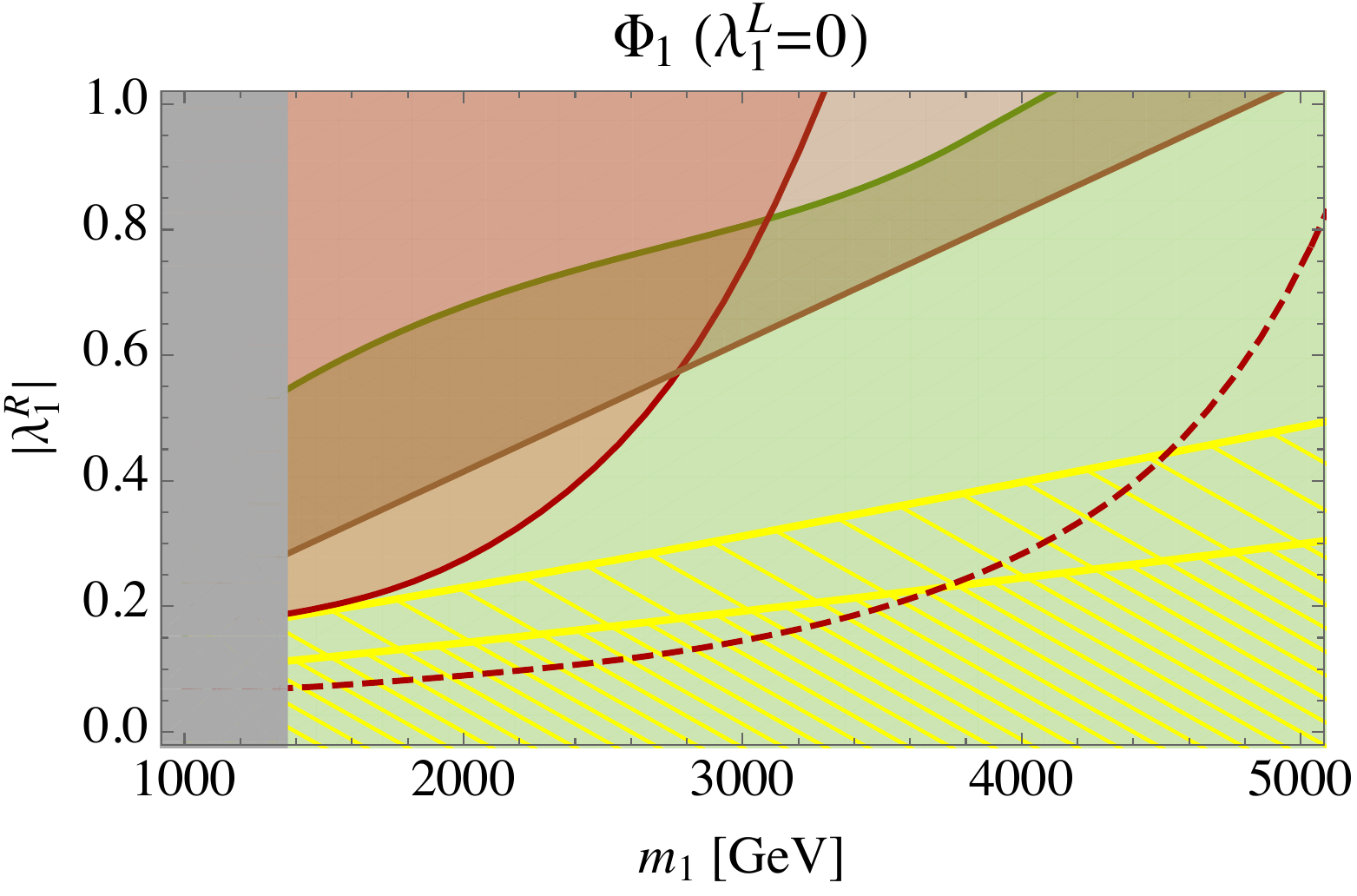}\\
		\includegraphics[width=0.45\textwidth]{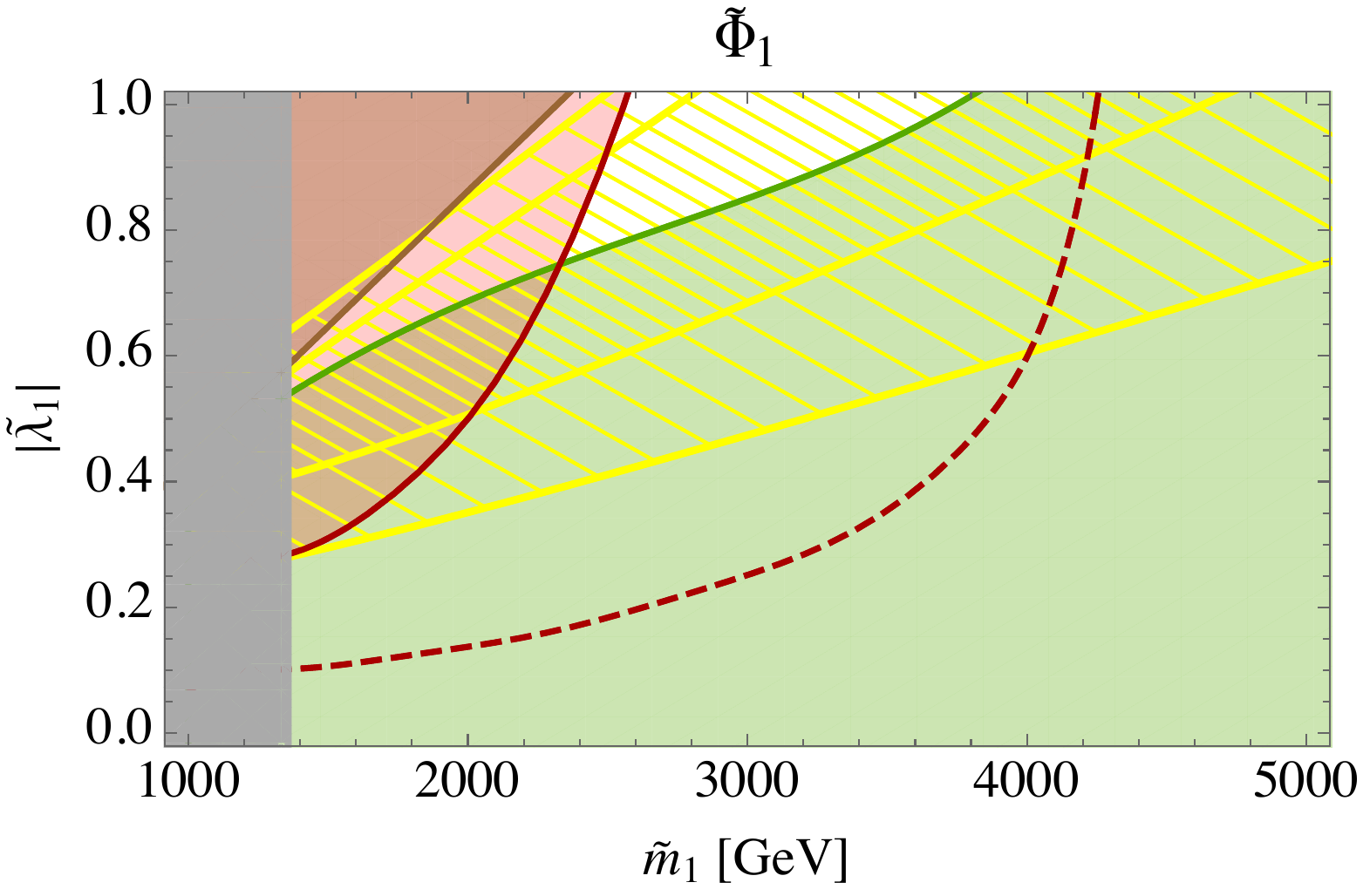}\qquad
		\includegraphics[width=0.45\textwidth]{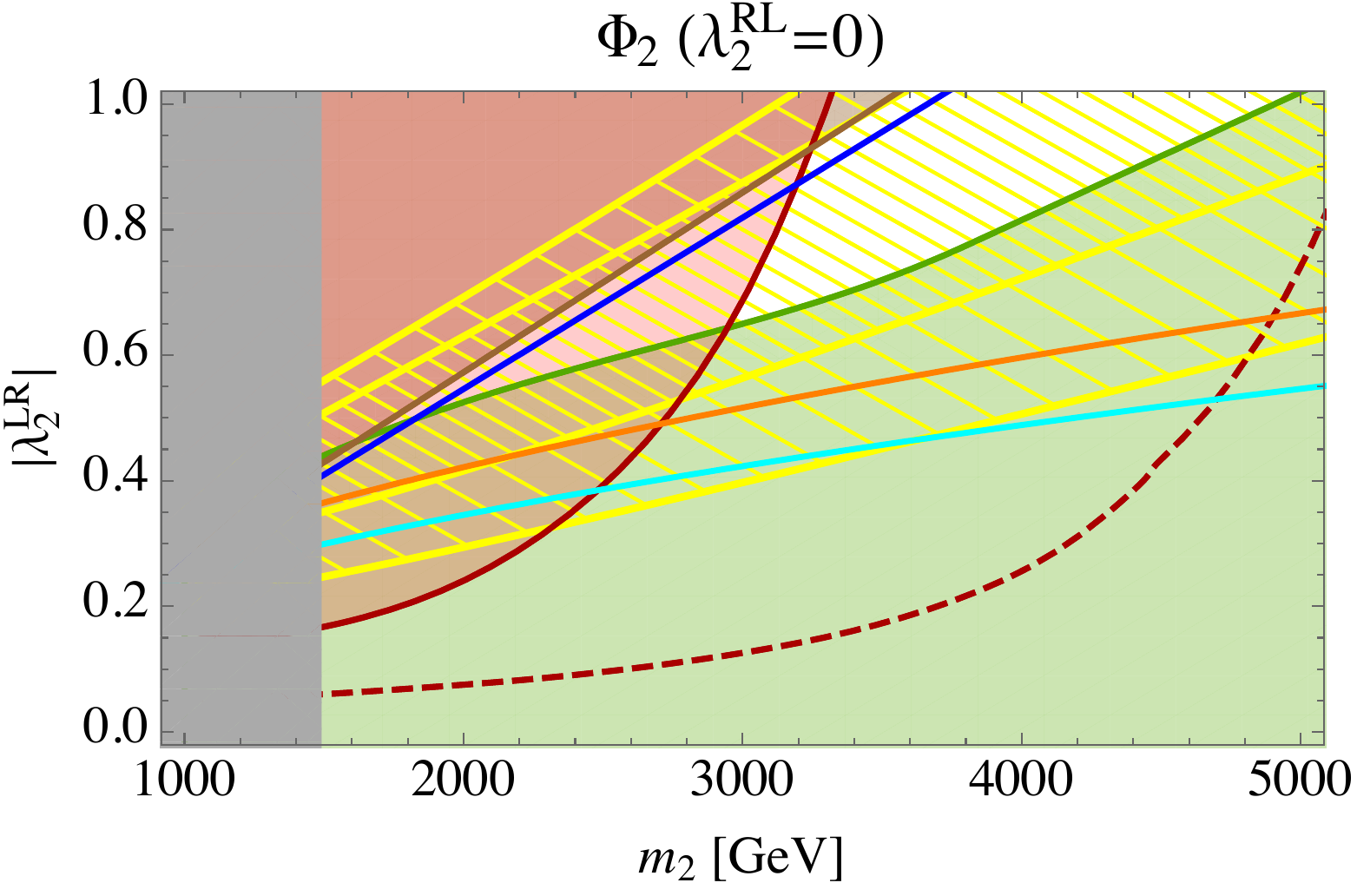}\\
		\includegraphics[width=0.45\textwidth]{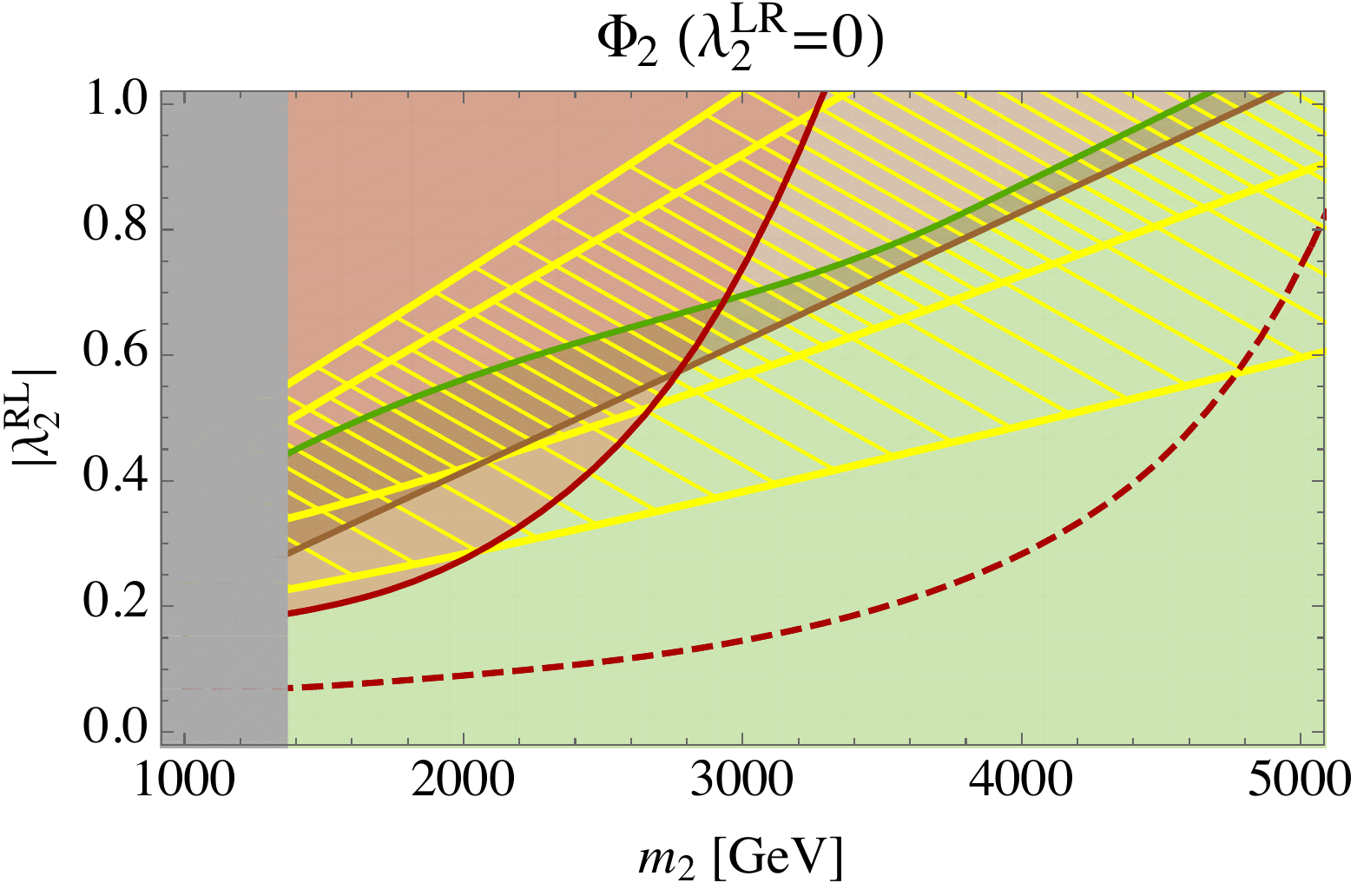}\qquad
		\includegraphics[width=0.45\textwidth]{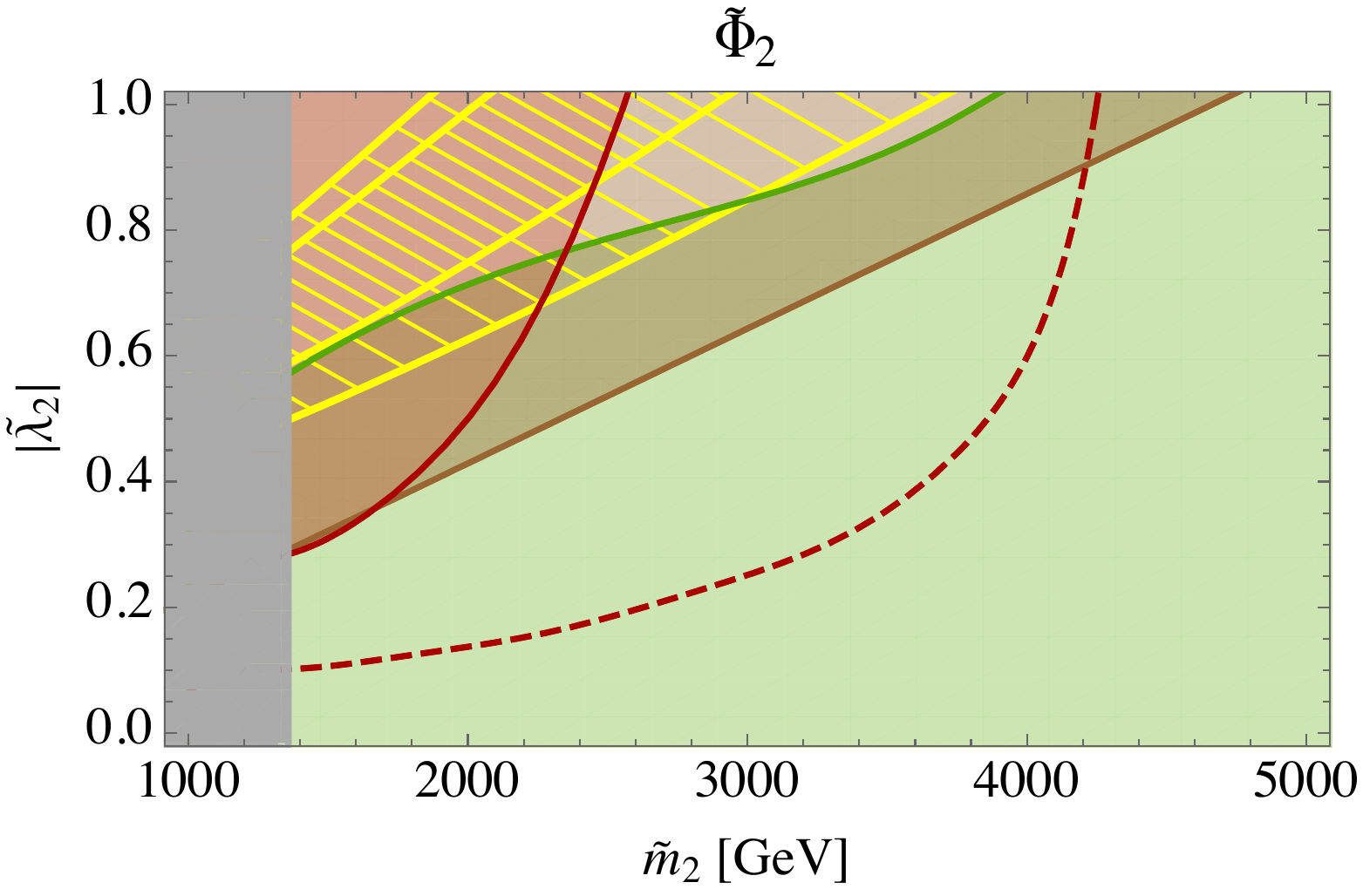}\\
		\includegraphics[width=0.9\textwidth]{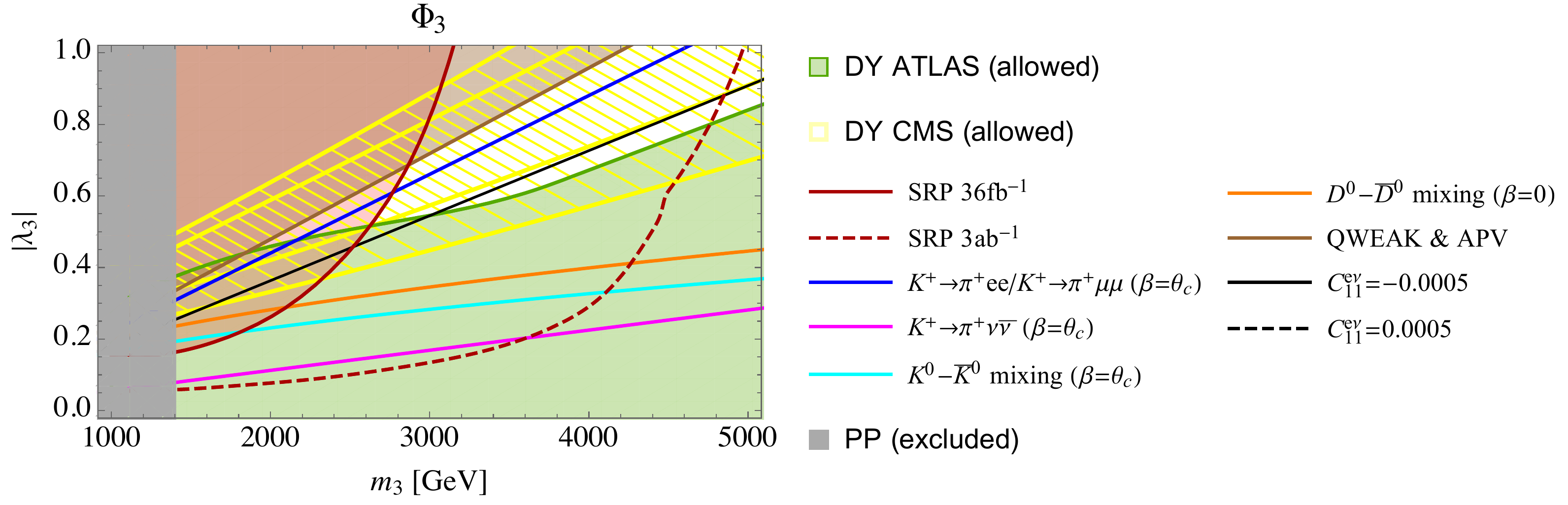}\qquad \qquad
	\caption{Limits on the parameter space of first generation scalar LQs. The regions above the colored lines are excluded. {The preferred $1\sigma$ and $2 \sigma$ regions by CMS measurement are shown in yellow}. While LHC limits and the bounds from parity violation are to a good approximation independent of {$\beta$ (for $\beta= O(\theta_c)$)} the bounds from kaon and $D$ decays depend on it. We consider the two scenarios $\beta=\theta_{c}$ or $\beta=0$. In the first case, the kaon limits arise for LQ representations with left-handed quark fields while in the second case these limits are absent but bounds from $D^{0}-\bar{D}^{0}$ arise.}
	\label{fig:SLQ_plot}
	\end{center}
\end{figure*}
\begin{figure*}
	\begin{center}
		\includegraphics[width=0.45\textwidth]{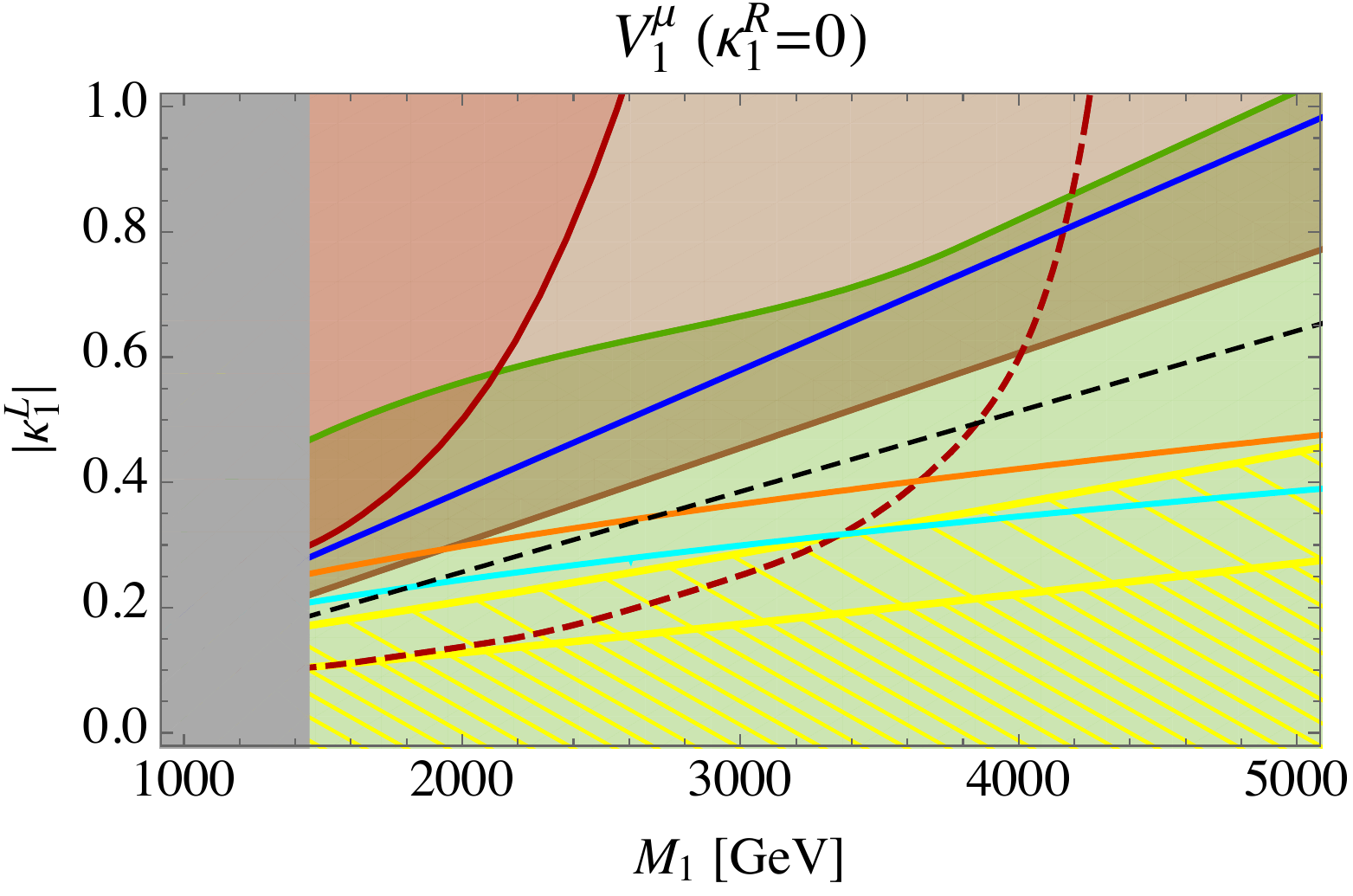}\quad
		\includegraphics[width=0.45\textwidth]{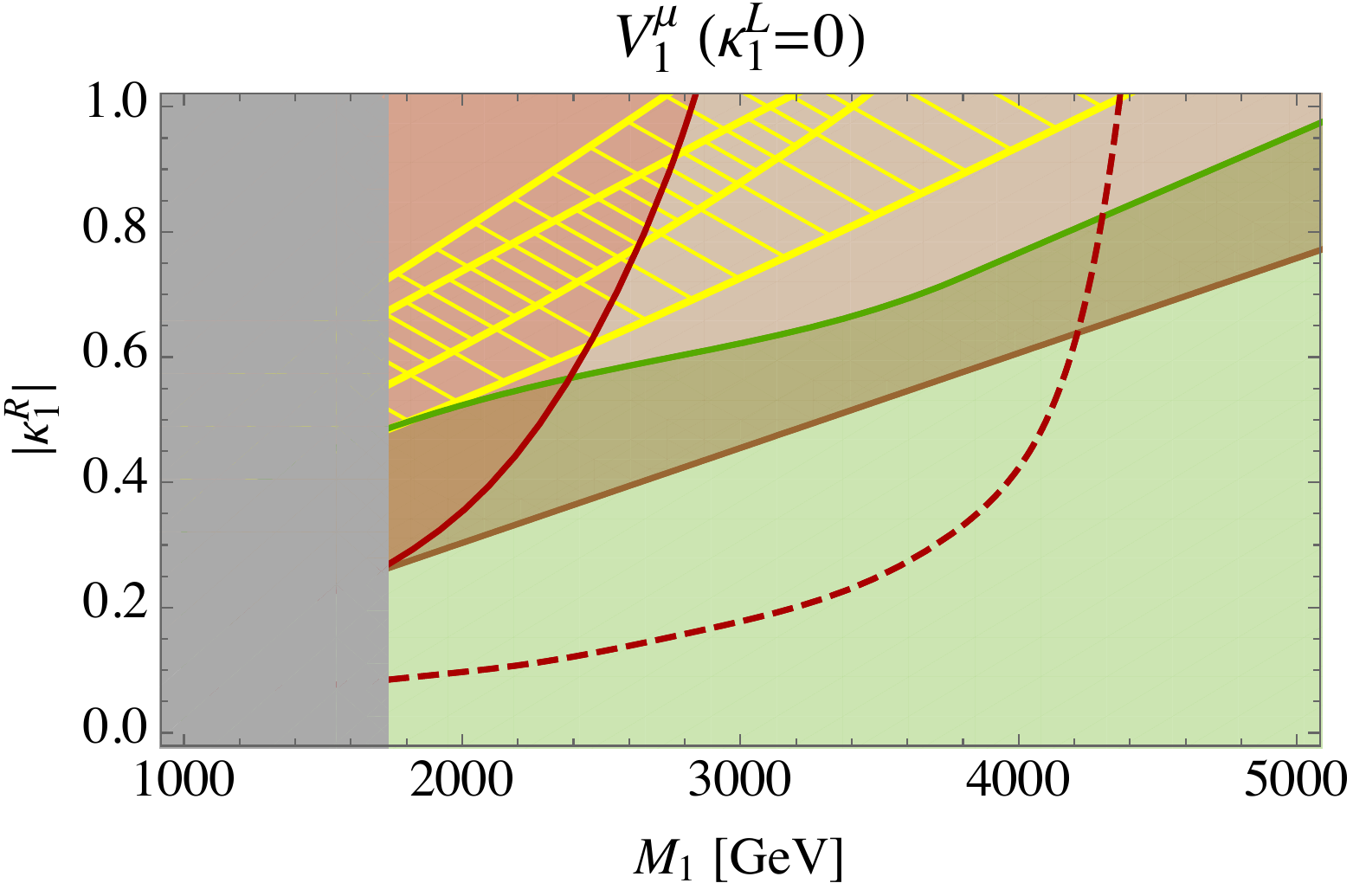}\\
		\includegraphics[width=0.45\textwidth]{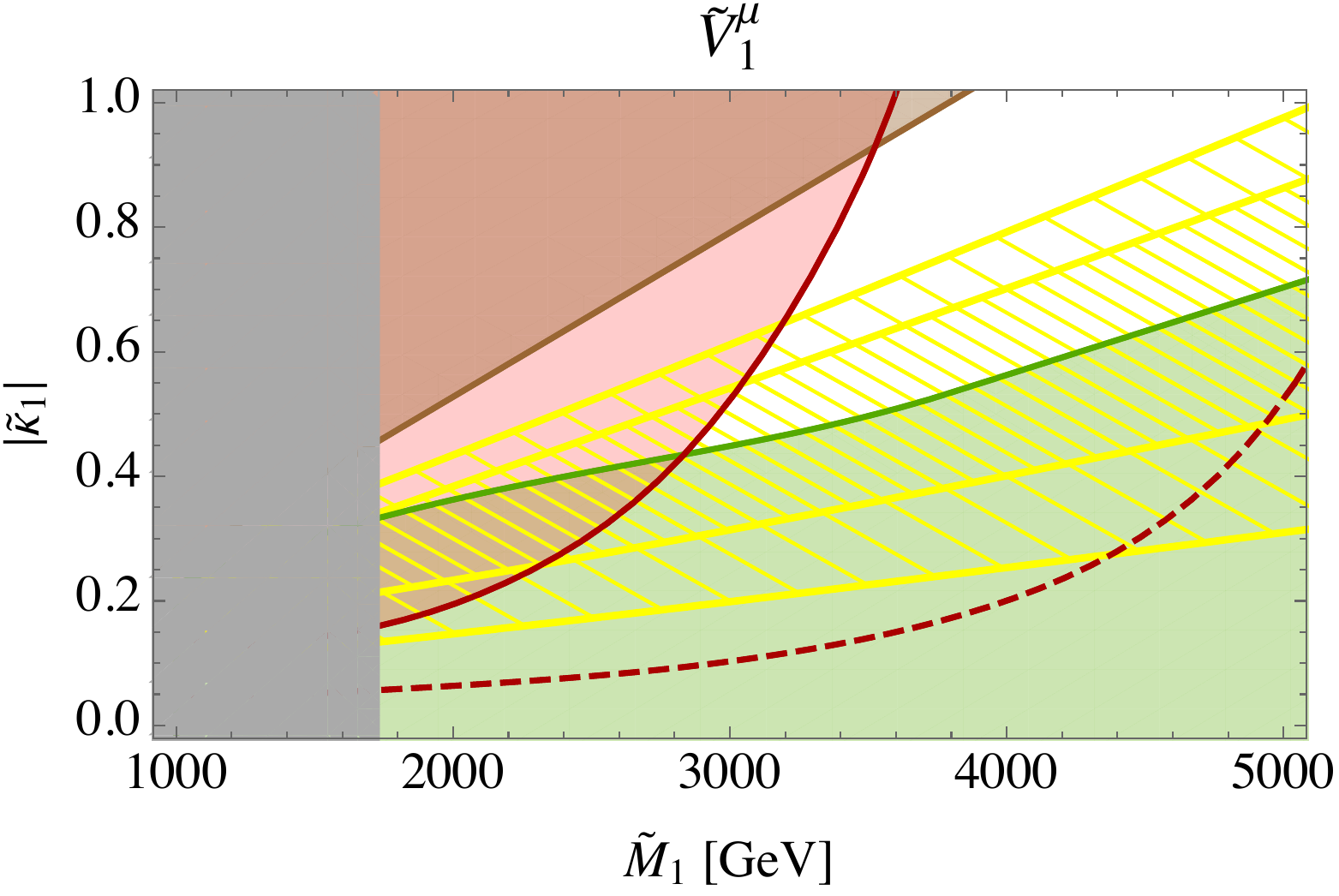}\qquad
		\includegraphics[width=0.45\textwidth]{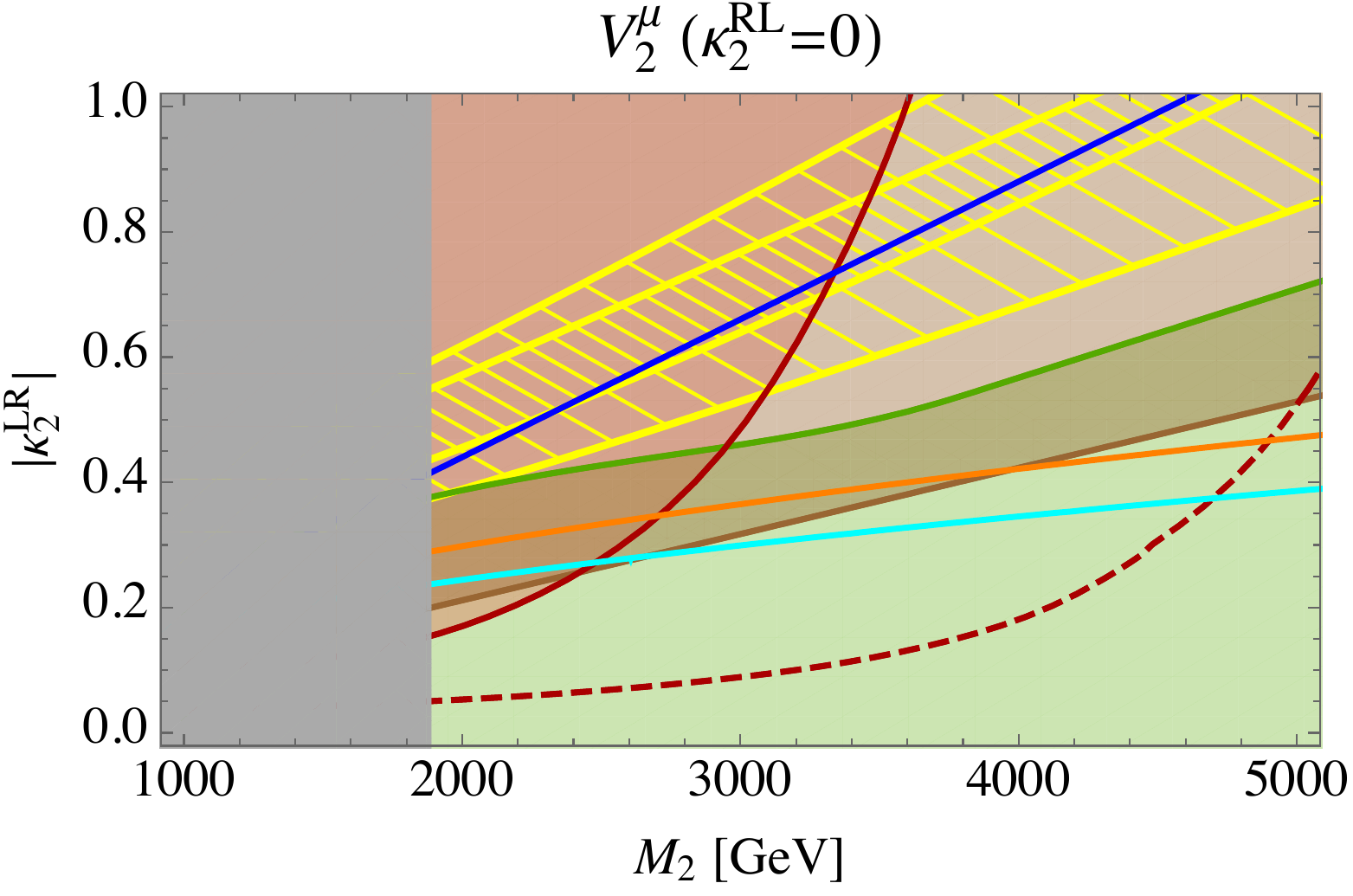}\\
		\includegraphics[width=0.45\textwidth]{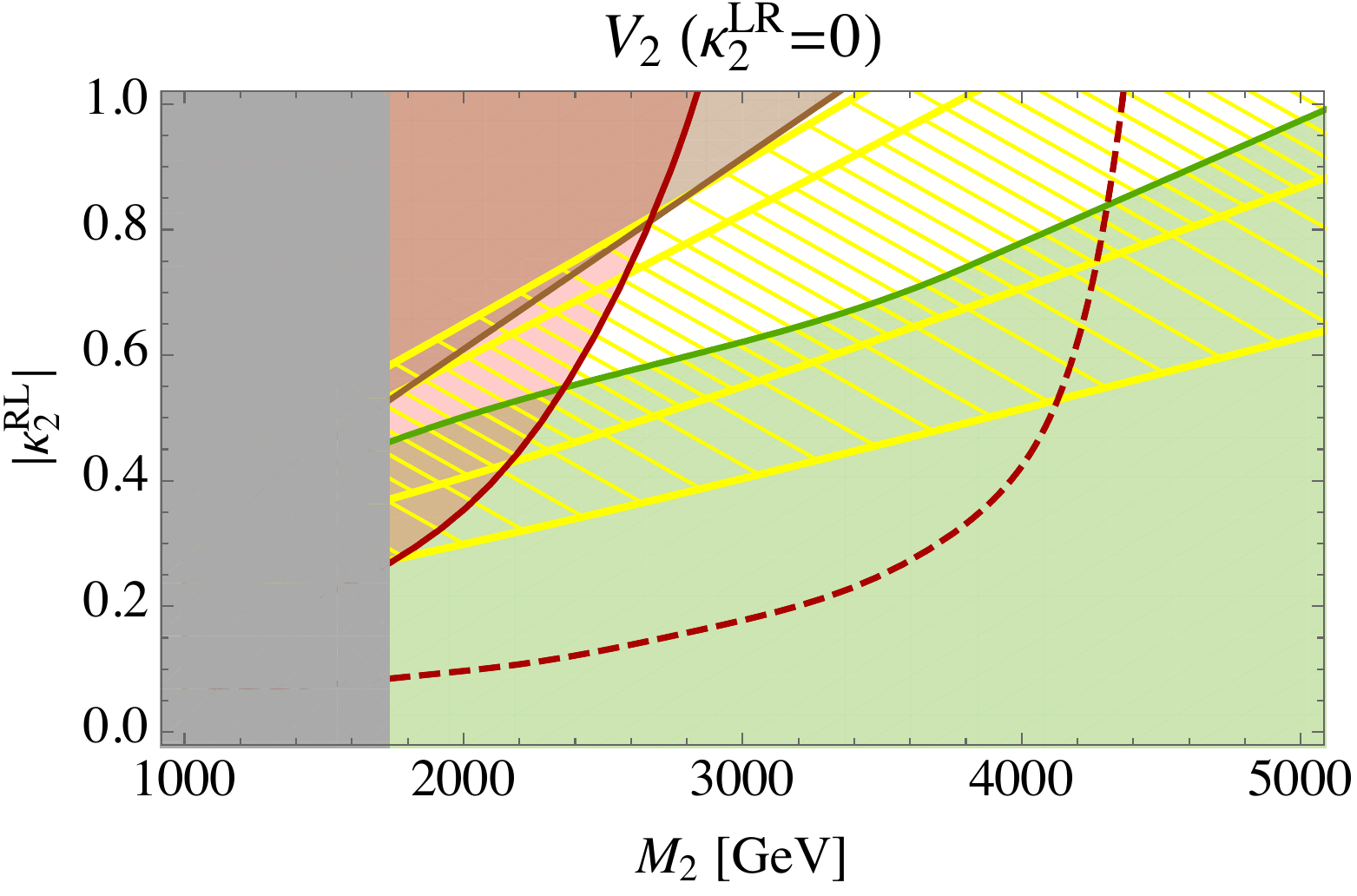}\qquad
		\includegraphics[width=0.45\textwidth]{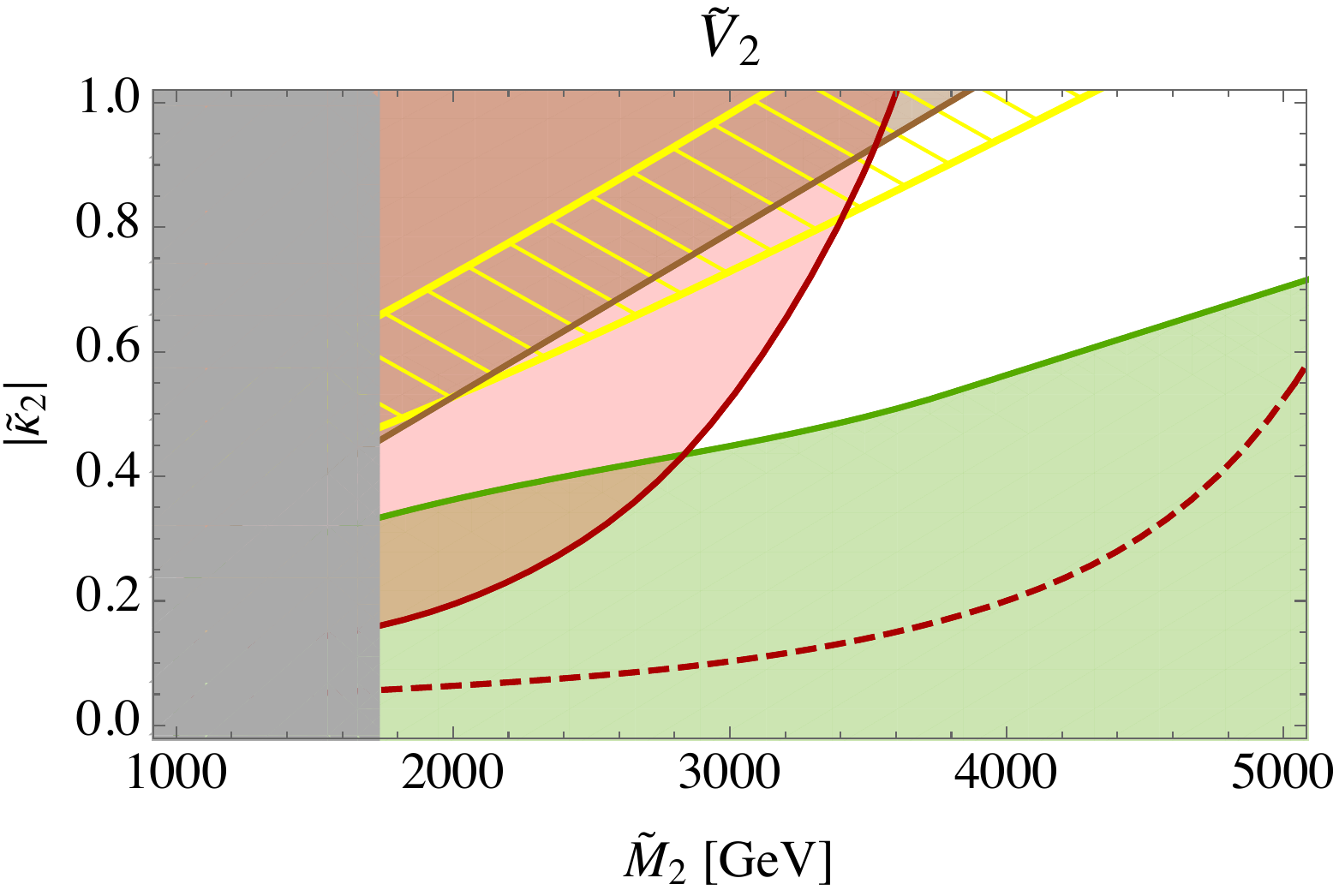}\\
		\includegraphics[width=0.9\textwidth]{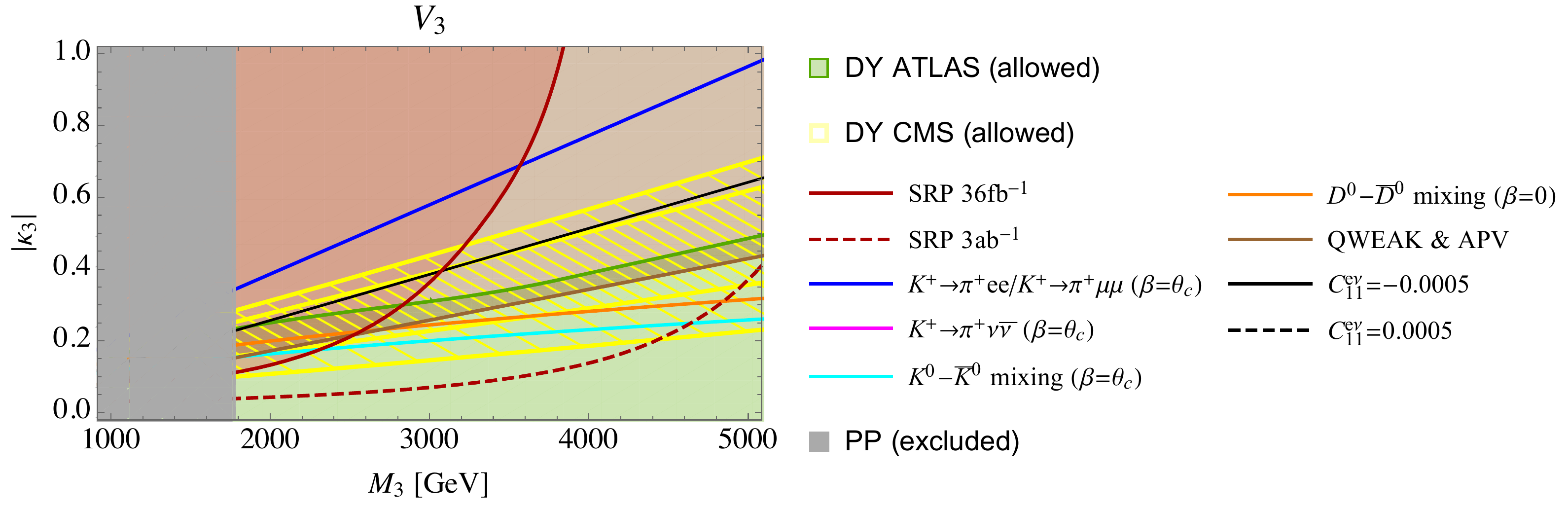}\qquad \qquad
	\caption{Limits on the parameter space of first generation vector LQs. The regions above the colored lines are excluded. {The $1\sigma$ and $2 \sigma$ regions that are preferred by the CMS measurement are shown in yellow}. While LHC limits and the bounds from parity violation are to a good approximation independent of {$\beta$ (for $\beta= O(\theta_c)$)} the bounds from kaon and $D$ decays depend on it. We consider the two scenarios $\beta=\theta_{c}$ or $\beta=0$. In the first case, the kaon limits arise for LQ representations with left-handed quark fields while in the second case these limits are absent but bounds from $D^{0}-\bar{D}^{0}$ arise.}
	\label{fig:VLQ_plot}
	\end{center}
\end{figure*}
\section{Phenomenological Analysis}
\label{sec:pheno}

The regions allowed at the $1\sigma$ and $2\sigma$ level (with respect to the best fit points) from the CMS measurement are shown in Fig.~\ref{fig:SLQ_plot} and \ref{fig:VLQ_plot} together with the other constraints on first generation LQs. For $\tilde{\lambda}_1, \lambda_2^{LR}, \lambda_2^{RL}, \tilde{\lambda}_2, \lambda_3, \kappa_1^R, \tilde{\kappa}_1, \kappa_2^{LR}, \kappa_2^{RL}, \tilde{\kappa}_2$ and $\kappa_3$, the model explains data better than the SM, with $\Delta \chi^2 \equiv \chi^2(\infty, 0) - \chi^2 (\hat{m}, \hat{\lambda})$ being $\approx-11$ as they provide clear effects in the high $m_{\ell \ell}$ bins. The representations with $\lambda_1^L, \lambda_1^R$ and $\kappa_1^L$ feature a destructive LQ--SM interference term. While they can also yield  $R_{\mu\mu/ee}^{SM+LQ}/R_{\mu\mu/ee}^{SM} < 1$ at large $m_{\ell \ell}$ values for sizable LQ contributions, the destructive interference then results in deviations $R_{\mu\mu/ee}^{SM+LQ}/R_{\mu\mu/ee}^{SM} > 1$ for intermediate invariant masses ($m_{\ell \ell} \approx 500$ GeV) leading to a fit that is worse than the one of the SM. The representations with couplings $\tilde{\lambda}_2, \kappa_1^R, \kappa_2^{LR}, \tilde{\kappa}_2$ also lead to destructive interference with the SM, but they feature smaller interference terms. In these cases, $R_{\mu\mu/ee}^{SM+LQ}/R_{\mu\mu/ee}^{SM}$ is only slightly larger than 1 for intermediate $m_{\ell \ell}$, allowing still for a good fit to the data.

Taking into account the other exclusion limits already presented in Ref.~\cite{Crivellin:2021egp}, we see from Fig.~\ref{fig:SLQ_plot} and \ref{fig:VLQ_plot} that the representations with the couplings  $\tilde{\lambda}_1$, $\lambda_2^{LR}$, $\lambda_2^{RL}$, $\tilde{\kappa}_1$, $\kappa_2^{RL}$ and $\kappa_3$ can account for the CMS measurement without violating other bounds, since these representations interfere constructively with the SM such that small LQ contributions are sufficient to explain the excess in electron pairs. $\lambda_2^{LR}$ and $\kappa_3$ are potentially in slight tension with the neutral meson mixing or rare kaon decays, depending on the CKM mixing angle $\beta$. However, in case of alignment to the up sector, and allowing for fine-tuning in $D^0-\bar D^0$ mixing, $V_3$ could not only account for the CMS data, but also explain the Cabbibo angle anomaly.  Although the representations with the couplings $\tilde{\lambda}_2, \lambda_3, \kappa_1^R, \kappa_2^{LR}$ and $\tilde{\kappa}_2$ explain the CMS data, they are excluded by parity violation, $K^+ \to \pi^+ \nu \bar{\nu}$, or the ATLAS Drell-Yan measurement. Except for $\lambda_3$, these representations all interfere destructively with the SM, requiring excluded lower values of $\hat{m}/\hat{\lambda}$ (corresponding to larger LQ contributions) to explain the electron excess at high $m_{\ell \ell}$.

\medskip 

\section{Conclusion}
\label{sec:conclusion}
In this addendum we examined the impact of the CMS measurement on LFU violation in non-resonant di-lepton pairs. We found that the first generation LQs with the couplings $\tilde{\lambda}_1, \lambda_2^{LR}, \lambda_2^{RL}, \tilde{\lambda}_2, \lambda_3, \kappa_1^R, \tilde{\kappa}_1, \kappa_2^{LR}, \kappa_2^{RL}, \tilde{\kappa}_2$ and $\kappa_3$ provide better fits to the CMS data than the SM with $\Delta\chi^2 \approx -11$. Among these, $\tilde{\lambda}_1$, $\lambda_2^{LR}$, $\lambda_2^{RL}$, $\tilde{\kappa}_1$, $\kappa_2^{RL}$, $\kappa_3$ feature constructive interference with the SM and are consistent with all other available measurements. In case of alignment to the up sector and allowing for fine-tuning in $D^0-\bar{D}^0$ mixing, $V_3$ could not only account for the CMS data, but also explain the Cabbibo angle anomaly.\\
\medskip
\begin{acknowledgements}
A.C. is grateful to Emanuele Bagnaschi for bringing the CMS analysis to his attention. The work of A.C. is supported by a Professorship Grant (PP00P2\_176884) of the Swiss National Science Foundation. L.S. is supported by the ``Excellence Scholarship \& Opportunity Programme'' of the ETH Z\"urich. A.C. thanks CERN for the support via the Scientific Associate program.
\end{acknowledgements}

\bibliography{LQ1addendum}

\end{document}